\documentclass[prb,twocolumn,showpacs,superscriptaddress]{revtex4}
\usepackage{graphicx}
\usepackage{amsmath}
\usepackage{amssymb}

\renewcommand{\i}{\mathrm{i}}
\newcommand{\e}{\mathrm{e}}
\newcommand{\p}{\partial}

\newcommand{\bsigma}{\boldsymbol{\sigma}}
\DeclareMathAlphabet{\bi}{OML}{cmm}{b}{it}

\begin{document}
\title{Enhanced Weiss oscillations in graphene}
\author{A. Matulis}\email{amatu@pfi.lt}
\affiliation{Departement Fysica, Universiteit Antwerpen \\
Groenenborgerlaan, B-2020 Antwerpen, Belgium}
\affiliation{Semiconductor Physics Institute, Go\v{s}tauto 11,
LT-01108 Vilnius, Lithuania}
\author{F.~M.~Peeters }
\email{francois.peeters@ua.ac.be}
\affiliation{Departement Fysica, Universiteit Antwerpen \\
Groenenborgerlaan, B-2020 Antwerpen, Belgium}

\begin{abstract}
The magneto-conductivity of a single graphene
layer where the electrons are described by the Dirac Hamiltonian
weakly modulated by a periodic potential is calculated.
It is shown that Weiss oscillations periodic in the inverse
magnetic field appear, that are more pronounced
and less damped with the increment of temperature as
compared with the same oscillations in a typical two-dimensional
electron system with a standard parabolic energy spectrum.
\end{abstract}


\date{\today}

\maketitle

\section{Introduction}

Recently the successful preparation of monolayer graphene films
\cite{nov06,zhang05,berger06} has generated a lot of activity in
the physics of two-dimensional (2D) Dirac fermions. The massless
energy spectrum and the specific density of states of electrons
and holes described by the Dirac Hamiltonian enabled to study
experimentally the chiral tunnelling and Klein paradox in graphene
\cite{kat06}, and led to the discovery of the unconventional
``half-integer quantum Hall effect'' \cite{nov05,zhang05,gus05}.
The presence of holes in graphene with 2D Dirac-like spectrum was
confirmed by measurements of de Haas-van Alphen and Shubnikov-de Haas
(SdH) oscillations. \cite{gus04} These magnetic oscillations appear
due to the interplay of the quantum Landau levels with the Fermi energy
in the metal, and serve as a powerful technique to investigate
the Fermi surface and the spectrum of electron excitations.

Another technique which was successfully used to obtain
information on the electron spectrum of 2D systems is based on the
interaction of electrons with artificially created periodic
potentials with periods in the submicrometer range. Such
electrical modulation of the 2D system was created by two
interfering laser beams \cite{weiss89}, or by depositing an array
of parallel metallic strips on the layer surface \cite{wink89},
and led to the discovery of Weiss oscillations in the
magneto-resistance. These oscillations are a consequence of the
commensurability of the electron cyclotron orbit radius at the
Fermi energy and the period of the above electrical modulation.
They were found to be periodic in the inverse magnetic field like
the SdH oscillations, but have a different period versus electron
density dependence. The period for Weiss oscillations varies with
electron density ($n_e$) as $\sqrt{n_e}$, whereas that of the SdH
ones as $n_e$. Theoretical calculations of these oscillations were
presented in Refs.~\cite{gerh89,wink89,vas89}, and it was shown
that Weiss oscillations in the magneto-resistance for motion
perpendicular to the oscillating potential can be understood as
being a semiclassical effect.\cite{been89,peet92}

The Klein paradox in graphene where Dirac electrons can penetrate
through potential barriers with a rather high probability shows that
electric control of Dirac electrons cannot be realized. Therefore,
it is interesting to investigate the sensitivity of Dirac
electrons on the electrical modulation of the layer.
Thus, we subjected the system to a periodic potential that introduces
a new length scale and a new energy scale into the problem.
We found that such a periodic potential also leads to Weiss oscillations
in graphene which are even more pronounced than in typically 2D
electron gases with a parabolic electron spectrum.

The paper is organized as follows. In the next Sec.~all necessary
expressions for the magneto-conductivity calculation are given, and in
Sec.~III the obtained results for graphene layer are compared
with results for the standard 2D electron system. In Sec.~IV
the asymptotic expression valid in the quasi-classical region
is obtained. In Sec.~V some simple classical explanation
of obtained results are presented, and in the last Sec.~the
short conclusions are given.

\section{Electrical magnetotransport}

We consider the graphene layer within the single electron
approximation where the low energy excitations are described
by the two-dimensional (2D) Dirac-like hamiltonian \cite{kat06}
\begin{equation}\label{ham0}
  H_0 = v_F\bsigma\left(-i\hbar\nabla + \frac{e}{c}\bf{A}\right).
\end{equation}
Here $\bsigma=\{\sigma_x,\sigma_y\}$ are the Pauli matrices, and
the vector potential ${\bf A}=\{0,Bx\}$ describing the magnetic
field ${\bf B}=\{0,0,B\}$ perpendicular to the graphene layer is
chosen in the Landau gauge. The parameter $v_F$ characterizes the
electron velocity which is usually about $300$ times smaller
than the velocity of light. The total Hamiltonian consists of
two parts
\begin{equation}\label{hamtot}
  H = H_0 + U(x),
\end{equation}
where the additional potential
\begin{equation}\label{potmod}
  U(x) = V_0\cos(2\pi x/a_0)
\end{equation}
describes the static electrical modulation of our 2D system
in the $x$ direction.

The conductivity tensor $\sigma_{\mu\nu}(\omega)$ within the
one-electron approximation was evaluated in Ref.~\cite{abe75}.
We shall restrict ourselves to the diffusion contribution
(i.~e.~which is the dominant contribution)
which stems from the diagonal part of the density operator.
Following Ref.~\cite{char82} it can be written as
\begin{equation}\label{cond0}
  \sigma_{yy} = \frac{\beta e^2}{L_xL_y}
  \sum_{\zeta} f(E_{\zeta})\left\{1 - f(E_{\zeta})\right\}
  \tau(E_{\zeta})\left(v_y^{\zeta}\right)^2.
\end{equation}
Symbols $L_x,L_y$ characterize
the dimensions of the layer, and $\beta$ is the inverse temperature.
This diagonal part is caused by the influence of the electrical
modulation on the electron drift in crossed electric and magnetic
fields and is in order of magnitude larger than the other diagonal
component $\sigma_{xx}$ which appears due to scattering on imperfections.
Symbol $\zeta$ denotes the quantum numbers of the electron eigenstates,
and the velocities $v_y^{\zeta}$ can be calculated as derivatives
of the energy eigenvalue over the corresponding electron momenta.

The resistivity tensor is given in terms of the inverse conductivity
tensor, namely, $\rho_{xx} = \sigma_{yy}/(\sigma_{xx}\sigma_{yy}-
\sigma_{xy}\sigma_{yx})$ what in our case of small conductivity
reduces to $\rho_{xx} \approx \sigma_{yy}/\sigma_{xy}^2$, and its
$\rho_{xx}$ component is actually proportional to the conductivity
in the considered perpendicular direction.

The energy eigenvalues are defined through the solution of the
stationary Schr\"{o}dinger equation
\begin{equation}\label{stsred}
  \{H - E\}\Psi({\bf r}) = 0
\end{equation}
with total Hamiltonian (\ref{hamtot}). We will follow
Ref.~\cite{peet92} and assume that the electrical modulation is
small enough and restrict to a lowest perturbation expansion in
$V_0$. For this purpose we have to solve the zero order
Schr\"{o}dinger equation with the Hamiltonian $H_0$. Due to the
system homogeneity along $y$ axis we substitute the eigenfunction
as
\begin{equation}\label{eigenfk}
  \Psi({\bf r}) = \frac{1}{\sqrt{L_y}}e^{ik_yy}\begin{pmatrix}
  a(x) \\ b(x) \end{pmatrix},
\end{equation}
and transform the zero order eigenvalue problem into the following two
ordinary differential equation set for the wave function components:
\begin{subequations}\label{srsetmg}
\begin{eqnarray}
\label{srsetmg1}
  -\i\hbar v_F\left(\frac{\p}{\p x} - \i \frac{\p}{\p y}
  + \frac{x}{l^2}\right)b - Ea &=& 0, \\
\label{srsetmg2}
  -\i\hbar v_F\left(\frac{\p}{\p x} + \i \frac{\p}{\p y}
  - \frac{x}{l^2}\right)a - Eb &=& 0,
\end{eqnarray}
\end{subequations}
where $l=\sqrt{c\hbar/eB}$ is the magnetic length.
Solution of these equations can be easily obtained making use of the
analogy with the harmonic oscillator eigenvalue problem. It reads
\begin{subequations}\label{dwf}
\begin{eqnarray}
\label{dwf1}
  E_n^{D} &=& \frac{v_F\hbar}{l}\sqrt{2n}, \\
\label{dwf2}
  \Psi_{n,k_y}({\bf r}) &=& \frac{\e^{\i k_yy}}{\sqrt{2L_yl}}\begin{pmatrix}
  -\i\Phi_{n-1}([x+x_0]/l) \\ \Phi_n([x+x_0]/l) \end{pmatrix},
\end{eqnarray}
\end{subequations}
where
\begin{equation}\label{ssol}
  \Phi_n(x) = \frac{\e^{-x^2/2}}{\sqrt{2^nn!\sqrt{\pi}}}
  H_n(x),
\end{equation}
is expressed in the normalized Hermitian polynomials, and
$x_0 = l^2k_y$ indicates the localization of the electron in
the $x$ direction.

For the first order correction one has to add to the energy
eigenvalue the diagonal matrix element of potential (\ref{potmod})
calculated with the above eigenfunctions:
\begin{equation}\label{encorr}
\begin{split}
  \Delta E_{n,k_y} &= \int_{-\infty}^{\infty}dx\int_0^{L_y}dy
  \Psi_{n,k_y}^+({\bf r})U(x)\Psi_{n,k_y}({\bf r}) \\
  &= \frac{V_0}{2}\cos(Kx_0)\e^{-u/2}
  \left\{L_n(u) + L_{n-1}(u)\right\},
\end{split}
\end{equation}
where $K=2\pi/a_0$, $u = K^2l^2/2$, and $L_n(u)$ is a Laguerre polynomial.
This energy correction makes the degenerate Landau levels $k_y$-dependent,
expands them into bands, what finally leads to non zero velocities:
\begin{equation}\label{veloc}
\begin{split}
  v_y^{\zeta} &\equiv v_y^{n,k_y}
  = \frac{1}{\hbar}\frac{\p}{\p k_y}\Delta E_{n,k_y} \\
  &= - \frac{V_0}{\hbar K}u\e^{-u/2}\left[
  L_n(u) + L_{n-1}(u)\right]\sin(Kx_0).
\end{split}
\end{equation}

Now substituting the velocities in Eq.~(\ref{cond0}) and specifying
the summation over quantum numbers as
\begin{equation}\label{summation}
  \sum_{\zeta} = \frac{L_y}{2\pi}\int_0^{L_x/l^2}dk_y\sum_{n=0}^{\infty}
\end{equation}
we obtain the final expression for the diffusion contribution for the
dc conductivity under the consideration
\begin{equation}\label{dcf}
  \sigma_{yy} = A\Phi,
\end{equation}
where
\begin{equation}\label{dcfa}
  A = 8\pi^2\frac{e^2}{h}\frac{V_0^2\tau\beta}{\hbar},
\end{equation}
and the function
\begin{equation}\label{dcfphi}
  \Phi = \frac{1}{4}u e^{-u}\sum_{n=0}^{\infty}\frac{f(E_n)}
  {\left[f(E_n) + 1\right]^2}
  \left[L_n(u) + L_{n-1}(u)\right]^2
\end{equation}
will be considered as a dimensionless conductivity.
In order to simplify this expression we introduced the following
exponential function $f(E)=\exp\left\{\beta(E-E_F)\right\}$
where $E_F$ is the Fermi energy.

\section{Comparison with the usual Weiss oscillations}

It is interesting to compare the obtained expression for dc
conductivity with the same conductivity calculated in Ref.~\cite{peet92}
for the case of a 2D electron system localized at the interface
between two semiconductors and having a parabolic energy
dependence on its momentum. Comparing velocity expression
(\ref{veloc}) with the equivalent one in Ref.~\cite{peet92} (see
Eq.~(4)) we notice that the diffusive conductivity for the system
of Dirac electrons differs from the one for the standard interface
case by replacement of the Laguerre polynomial $L_n(u)$ by the
average of two successive polynomials $\left[L_n(u) +
L_{n-1}(u)\right]/2$. And of course, two different expressions for
the Landau level energies (Eq.~(\ref{dwf1}) for Dirac electrons
and
\begin{equation}\label{enstd}
  E_n^P = \hbar\omega_c(n+1/2), \quad \omega_c = \frac{eB}{mc}
\end{equation}
for the parabolic electron spectrum case) has to be used.

\begin{figure}[h]
\begin{center}\leavevmode
\includegraphics[width=\linewidth]{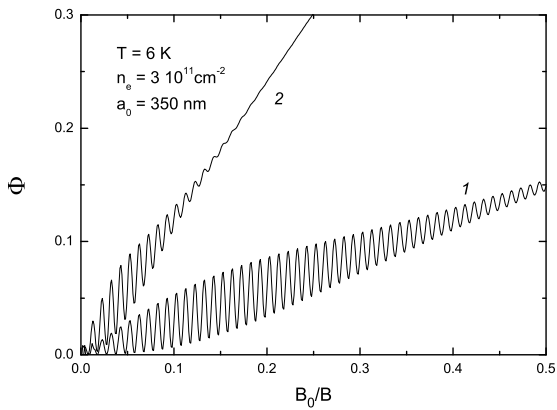}
\caption{The dimensionless conductivity versus inverse
magnetic field: {\it 1} -- Dirac electrons,
{\it 2} -- electrons with the parabolic energy spectrum.}
\label{fig1}
\end{center}
\end{figure}

These differences in the expressions, however, lead to essentially
different results for the dimensionless conductivity
as shown in Fig.~\ref{fig1}. The results are shown
as function of the inverse magnetic field for the temperature $T = 6\,$K,
electron density $n_e = 3\cdot 10^{11}$cm$^{-2}$, and the
period of electric modulation $a_0 = 350\,$nm.
The dimensionless magnetic field $b = B/B_0$ is introduced using
the characteristic
magnetic field $B_0 = c\hbar/ea_0^2$ corresponding to the magnetic
length equal to the modulation period $a_0$, which in the above
case is equal to $B_0 = 1$T.

We see that in graphene (curve {\it 1}) the Weiss oscillations
are more pronounced as compared with the system of electrons
with the standard parabolic energy spectrum (curve {\it 2}).
Furthermore, the Weiss oscillations in graphene are much more
robust with respect to temperature
damping in the quasi-classical region of small magnetic fields.
The physical reasons for these differences lay in different
Fermi velocities of Dirac and standard electrons. In order to
confirm the above statement we shall consider the asymptotic
behavior of Weiss oscillations in the quasi-classical region
which according to Refs.~\cite{gerh89,wink89} describes the
main features of the above oscillations and which also allow
for explicit analytical expressions.

\section{Asymptotic expressions}

The asymptotic expression for the conductivity (\ref{dcfphi})
is obtained following the approach of Ref.~\cite{peet92}
for case of standard electrons.
That approach is applicable when many Landau levels are filled, and
it is based on the following asymptotic expression for the
Laguerre polynomials:
\begin{equation}\label{lagass}
  \e^{-u/2}L_n(u) \to \frac{1}{\sqrt{\pi\sqrt{nu}}}
  \cos\left(2\sqrt{nu} - \pi/4\right).
\end{equation}
Taking the continuum limit
\begin{equation}\label{sumint}
  n \to \frac{1}{2}\left(\frac{lE}{v_F\hbar}\right)^2, \quad
  \sum_{n=0}^{\infty} \to \left(\frac{l}{v_F\hbar}\right)^2
  \int_0^{\infty}EdE,
\end{equation}
and having in mind that $u=2\pi^2/b$,
we transform Eq.~(\ref{dcfphi}) into the following integral:
\begin{equation}\label{quasicl0}
\begin{split}
  \Phi =& \frac{\sqrt{u}}{\pi}\left(\frac{l}{v_F\hbar}\right)^2
  \int_0^{\infty}\frac{f(E)EdE}{[f(E)+1]^2\sqrt{n}} \\
  &\times \cos^2\left(\frac{1}{2}\sqrt{\frac{u}{n}}\right)
  \cos^2\left(2\sqrt{nu} - \frac{\pi}{4}\right) \\
  =& \frac{2a_0}{v_F\hbar b}
  \int_0^{\infty}\frac{f(E)dE}{[f(E)+1]^2}
  \cos^2\left(\frac{\pi v_F\hbar}{Ea_0}\right) \\
  &\times \cos^2\left(\frac{2\pi a_0E}{v_Fb\hbar}
  - \frac{\pi}{4}\right).
\end{split}
\end{equation}

Now assuming that the temperature is low ($\beta^{-1} \ll E_F$)
and replacing $E = E_F + s\beta^{-1}$ we rewrite the above integral
as
\begin{equation}\label{quasicl1}
\begin{split}
  \Phi =& \frac{2a_0}{v_F\hbar b\beta}
  \cos^2\left(\frac{\pi}{p}\right)
  \int_{-\infty}^{\infty}\frac{ds\, e^s}{(e^s+1)^2} \\
  &\times \cos^2\left(\frac{2\pi p}{b} - \frac{\pi}{4}
  + \frac{2\pi p}{b\beta}s\right),
\end{split}
\end{equation}
where symbol
\begin{equation}\label{fdimless}
  p = \frac{E_Fa_0}{v_F\hbar} = k_Fa_0 = \sqrt{2\pi n_e}\,a_0
\end{equation}
stands for the dimensionless Fermi momentum of the electron.
Note in Eq.~(\ref{quasicl1}) we replaced all energies $E$ by the
Fermi energy $E_F$ except that one which is included in the
last cosine function, where the small energy correction can
influence the damping of the Weiss oscillations.

The obtained expression for the dimensionless conductivity
can be calculated using the standard integral
\begin{equation}\label{intstd}
  \int_0^{\infty}\frac{\cos(ax)}{\cosh(\beta x)}dx
  = \frac{\pi}{2\beta\sinh(a\pi/2\beta)},
\end{equation}
and the result can be presented as
\begin{equation}\label{result}
\begin{split}
  \Phi =& \frac{T}{4\pi^2 T_D}
  \cos^2\left(\frac{\pi}{p}\right)\Bigg\{\left[1
  - A\left(\frac{T}{T_D}\right)\right] \\
  &+ 2A\left(\frac{T}{T_D}\right)
  \cos^2\left[2\pi\left(\frac{p}{b}
  - \frac{1}{8}\right)\right]\Bigg\},
\end{split}
\end{equation}
where
\begin{equation}\label{afk}
  A(x) = \frac{x}{\sinh(x)} \stackrel{x\to\infty}{\to} = 2xe^{-x},
\end{equation}
and the symbol $T_D$ defined as
\begin{equation}\label{crtemp}
  k_BT_D = \frac{b\hbar}{4\pi^2 a_0}v_F
\end{equation}
($k_B$ is the Boltzman constant) gives the temperature scale
 for damping of the Weiss oscillations.

\begin{figure}[h]
\begin{center}\leavevmode
\includegraphics[width=\linewidth]{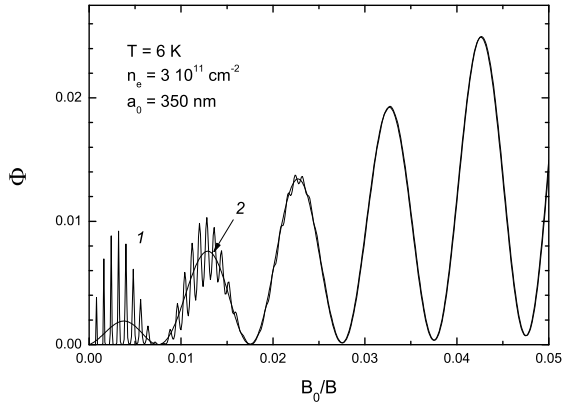}
\caption{The graphene conductivity as a function of
the inverse magnetic field: {\it 1} -- exact solution
(\ref{dcfphi}),
{\it 2} -- asymptotic expression (\ref{result}).}
\label{fig2}
\end{center}
\end{figure}

The validity of the asymptotic expression is seen in Fig.~\ref{fig2}
where the Weiss oscillations for the Dirac electrons (curve {\it 1})
are shown together with their asymptotic expression (curve {\it{2}})
for the same parameter values as in Fig.~\ref{fig1} in the region
of strong magnetic fields where deviations of the asymptotic
expression to the exact expression are largest.
We see that the coincidence of the exact result and its
asymptotic expression is rather good everywhere except
the region of very strong magnetic field where SdH oscillations
become superimposed on top of the Weiss oscillations.

We compare now the obtained result for the conductivity for the
Dirac electron system with the similar asymptotic result for the system
of electrons with the standard parabolic energy spectrum taken
from Ref.~\cite{peet92} which can be presented as follows:
\begin{equation}\label{dcndf}
\begin{split}
  \Phi =& \frac{T}{4\pi^2T_P}\Bigg\{\left[1
  - A\left(\frac{T}{T_P}\right)\right] \\
  &+ 2A\left(\frac{T}{T_P}\right)
  \cos^2\left[2\pi\left(\frac{p}{b}
  - \frac{1}{8}\right)\right]\Bigg\},
\end{split}
\end{equation}
where the critical temperature reads
\begin{equation}\label{tp}
  k_BT_P = \frac{bp\hbar^2}{4\pi^2ma_0^2}.
\end{equation}
Having in mind that
\begin{equation}\label{ftr}
  \frac{p}{m} = \frac{k_Fa_0}{m} = \frac{a_0}{\hbar}v_F^P,
\end{equation}
where $v_F^P$ is the velocity of the standard electron on
the Fermi surface, the critical temperature can be presented as
\begin{equation}\label{tp0}
  k_BT_P = \frac{b\hbar}{4\pi^2a_0}v_F^P.
\end{equation}
For parameters used in Fig.~\ref{fig1} plot $k_F \sim 1.4\cdot 10^6$,
$p \sim 50$, and $\cos(\pi/p)\sim 1$. Thus the asymptotic
behavior of the dimensionless conductivity for Dirac electrons
(\ref{result}) and standard electrons with parabolic energy
spectrum (\ref{dcndf}) differ mostly due to the very different
critical temperatures.

Comparing Eqs.~(\ref{crtemp}) and (\ref{tp0}) we see that
\begin{equation}\label{tratio}
  \frac{T_P}{T_D} = \frac{v_F^P}{v_F},
\end{equation}
or the ratio of critical temperatures for the standard
and Dirac electrons is equal to the ratio of the corresponding
velocities on the Fermi surface. It can be estimated as
\begin{equation}\label{tratio1}
  \frac{T_P}{T_D} = \frac{\hbar\sqrt{2\pi n_e}}{v_Fma_0}
\end{equation}
and for typical cases is less than unity. For instance, in
the case of the parameters used in Fig.~\ref{fig1} it
is $T_P/T_D \sim 0.24$ what explains the different slope and
damping of the Weiss oscillations.

\section{Quasi-classical explanation}

The obtained results can be understood from a simple physical picture.
In order to estimate the oscillation period we write down the
momentum of the electrons on the Fermi surface:
\begin{equation}\label{fmom}
  p_F = m\omega_cR_c
\end{equation}
which is identical in both cases. Then it follows that the
radius of electron orbit in the magnetic field is
\begin{equation}\label{radius}
  R_c = \frac{p_F}{m\omega_c} = \frac{\hbar k_Fc}{eB}
  = \frac{p l^2}{a_0}.
\end{equation}
The physical reason for the appearance of the Weiss oscillations is
the commensurability of the electron orbit radius with the period
of the electrical modulation, consequently, the argument of the
cosine function has to be proportional to
\begin{equation}\label{freq}
  \frac{R_c}{a_0} = p\frac{l^2}{a_0^2} = \frac{p}{b},
\end{equation}
which we can see in both expressions (\ref{result}) and (\ref{ftr}).

The damping factor of the oscillations can be estimated as follows.
Due to the finite temperature there are electron orbits with various
radia. The effective damping can be estimated as the ratio of
the above broadening of the orbit to the period of the modulation:
\begin{equation}\label{smear}
  \delta = \frac{1}{a_0}\frac{\p R_c}{\p E_F}\beta^{-1}.
\end{equation}

In the case of standard electrons (when $E_F=p_F^2/2m$) it reads
\begin{equation}\label{broadstd}
\begin{split}
  \delta_P &= \frac{m}{a_0\beta p_F}\frac{\p R_c}{\p p_F}
  = \frac{m}{a_0\beta\hbar k_Fm\omega_c} \\
  &= \frac{k_BT}{p\hbar\omega_c}
  = \frac{ma_0^2}{\hbar^2p b}k_BT,
\end{split}
\end{equation}
in agreement with the definition of the critical temperature
for standard electron (\ref{tp}).

In the case of Dirac electrons (when $E_F = v_Fp_F$)
the above damping parameter can be estimated as
\begin{equation}\label{broaddir}
  \delta_D = \frac{1}{\beta a_0v_F}\frac{\p R_c}{\p p_F}
  = \frac{ck_BT}{a_0v_FeB} = \frac{a_0}{\hbar b v_F}k_BT,
\end{equation}
what coincides with the obtained earlier result for
the critical temperature up to the same
constant $4\pi^2$. Now dividing Eq.~(\ref{broadstd})
by Eq.~(\ref{broaddir}) we obtain the same critical
temperature ratio dependence on the ratio of the Fermi
velocities (\ref{tratio}) what confirms the quasi-classical
nature of Weiss oscillations.

\section{Conclusions}

In conclusion, we studied the Weiss oscillations in electrically
modulated single layer of graphene. It was shown that the
static conductivity
oscillations are periodic in $1/B$ with period varying with
electron density as $\sqrt{n_e}$ like in the 2D electron system
with standard parabolic energy spectrum. Due to the larger
Fermi velocity the conductivity oscillations of the Dirac electron
system are more pronounced and less damped as compared with the 2D
system of electrons with parabolic energy spectrum in case
of analogous parameters. The found Dirac electron
sensitivity to the electric perturbation does not contradicts
the Klein paradox, because in contrary to electron tunneling
through barriers where both the electron and the hole nature of the
excitation plays a role, only the electrons at the Fermi energy are
responsible for the conductivity and for the studied Weiss
oscillations. Thus, their behavior in the electric field is not
overshadowed by the admixture of the hole states.

\begin{acknowledgments}
Part of this work is supported by the Flemish science foundation
(FWO-Vl) and the Belgian Science policy (PAI).
\end{acknowledgments}

\end{document}